\documentclass[conference,a4paper]{APSIPA2021}
\usepackage{amsmath}
\usepackage{graphicx}
\usepackage{multirow}
\usepackage{threeparttable}
\usepackage{amsmath}
\usepackage[psamsfonts]{amssymb}
\usepackage{amsxtra}
\usepackage{threeparttable}
\usepackage{comment}
\usepackage{multirow}
\usepackage{url}
\usepackage{mathrsfs}
\newcommand{\argmin}{\mathop{\rm arg~min}\limits}

\usepackage{geometry}
\usepackage{fancyhdr}

\fancypagestyle{firststyle}{
  \fancyhf{}
  \fancyhead[C]{2023 Asia Pacific Signal and Information Processing Association Annual Summit and Conference (APSIPA ASC)}
}

\begin{document}

\title{Toward Leveraging Pre-Trained Self-Supervised Frontends for Automatic Singing Voice Understanding Tasks: Three Case Studies}

\author{

Yuya Yamamoto\\
University of Tsukuba, Tsukuba, Japan \\
E-mail:s2130507@u.tsukuba.ac.jp}

\maketitle
\thispagestyle{firststyle}
\pagestyle{fancy}

\begin{abstract}
Automatic singing voice understanding tasks, such as singer identification, singing voice transcription, and singing technique classification, benefit from data-driven approaches that utilize deep learning techniques. 
These approaches work well even under the rich diversity of vocal and noisy samples owing to their representation ability. 
However, the limited availability of labeled data remains a significant obstacle to achieving satisfactory performance.
In recent years, self-supervised learning models (SSL models) have been trained using large amounts of unlabeled data in the field of speech processing and music classification. 
By fine-tuning these models for the target tasks, comparable performance to conventional supervised learning can be achieved with limited training data. 
Therefore, in this paper, we investigate the effectiveness of SSL models for various singing voice recognition tasks. 
We report the results of experiments comparing SSL models for three different tasks (i.e., singer identification, singing voice transcription, and singing technique classification) as initial exploration and aim to discuss these findings. 
Experimental results show that each SSL model achieves comparable performance and sometimes outperforms compared to state-of-the-art methods on each task.
We also conducted a layer-wise analysis to further understand the behavior of the SSL models.
\end{abstract}

\section{Introduction}
The singing voice plays an important role in the music. 
It provides emotional expressions for us through its melody and lyrics.
Computational understanding tasks of singing voices, such as singer identification, singing voice transcription, and singing expression identification, are beneficial for many applications such as music discovery \cite{ghias1995query}, pedagogy \cite{nakano2007mirusinger}, musicological analysis \cite{panteli2017towards}, etc.
Processing the singing voices is a long-running challenge in the music information retrieval (MIR) field \cite{humphrey2018introduction} due to its wide variation. 
Recently, the methods based on deep learning outperformed the conventional methods based on the hand-crafted manner in many singing voice understanding tasks thanks to its intense expressiveness \cite{gupta2022deep, Fu2019hierarchical,nasrullah2019music,yamamoto2021investigating}.
However, deep learning approaches typically necessitate extensive datasets comprising sung tracks with high-quality labels, entailing substantial costs for both data collection and annotation.

Transfer learning is one of the techniques to alleviate the requirements of large-scale datasets for the low-resource situation. 
It is based on the transfer of the knowledge derived from high-resource upstream pre-training tasks to target downstream tasks.
In the audio domain, there are many works on audio classification tasks that utilize transfer learning of PANNs \cite{kong2020panns} and VGGish \cite{hershey2017cnn}, which are pre-trained on a large-scale audio dataset.
Notably, transfer learning of the model that is pre-trained by self-supervised learning (SSL) fashion is rapidly emerging.
SSL models are leveraging a vast amount of unlabeled data, several notable models have been developed in both the speech and music domains.
In the speech domain, these models include Wav2Vec2.0 \cite{baevski2020wav2vec}, HuBERT \cite{hsu2021hubert}, and WavLM \cite{chen2022wavlm}. Meanwhile, in the music domain, models such as MERT \cite{li2023mert}, and MapMusic2Vec \cite{li2022map} have been introduced.

Singing voice encompasses characteristics of both speech and music, and researchers have explored leveraging pre-trained SSL models in singing voice understanding tasks \cite{ou2022transfer,heydari2022singing,gu2023deep,donahue2022melody} for each.
The transfer learning of pre-trained SSL models from the speech or music domains to the singing domain holds potential. 
However, there is still ample room for investigating the utility of each pre-trained SSL model.
This includes investigating which domains can contribute to singing voice understanding tasks, how the model extracts valuable features for the target tasks, and how to fully utilize the potential of the SSL models, and so on.


In this study, our objective is to examine the usefulness of SSL models pre-trained on the speech or music domains for singing voice understanding tasks. To achieve this, we employ pre-trained SSL models as a front-end to our voice identification model and evaluate their performance through fine-tuning.

We present the following contributions in this study:
\begin{enumerate}
    \item Comparative analysis of SSL models: We compare multiple SSL models across three distinct three tasks: singer identification, singing voice transcription, and singing technique classification, corresponding to "Who sings?", "What song?", and "How to sing?", respectively.
    \item Comparison with SoTA models: We also compared the SSL models with state-of-the-art (SoTA) models. Through fine-tuning the pre-trained SSL models, we demonstrate that they achieve performance comparable to that of current state-of-the-art methods in several singing voice understanding tasks.
    \item Investigation of layer-wise behavior: Additionally, we delve deeper into the behavior of the model's layers and analyze their characteristics across different tasks, by utilizing learnable weight for each layer. 
\end{enumerate}


\section{Related works}
Various machine learning approaches have been explored for low-resource problems in the singing voice understanding tasks.
Semi-supervised learning \cite{kum2020semi, hsu2021vocano}, data augmentation \cite{schluter2015exploring, hsieh2020addressing}, and self-supervised learning on singing voices \cite{yakura2022self, noufi2020self}, leveraging speech data \cite{basak2021end, zhang2022pdaugment} etc. have been proposed to mitigate the drawbacks of supervised-learning fashion.

More recent, pre-trained self-supervised models are used for singing understanding tasks.
Ou et al. leveraged Wav2Vec 2.0 \cite{baevski2020wav2vec} model that is pre-trained 
 using Librispeech corpus \cite{vassil2015librispeech} and fine-tuned automatic speech recognition for lyric transcription. They achieved comparable performance with the state-of-the-art performance of lyric transcription methods \cite{demirel2020automatic,ahlback2021mstre}, which are learned from over 150 hours of data, by using its 10\% of amount (i.e., 15 hours) \cite{ou2022transfer}.
 Gu et al. leveraged Wav2Vec2.0 for singing voice transcription and outperformed conventional works \cite{gu2023deep}.
 Heydari et al. tackled singing beat tracking, which is a beat tracking method when the input is only a singing voice. They leverage WavLM \cite{chen2022wavlm} and DistilHuBERT \cite{chang2022distilhubert} for the frontend of the model, and they outperformed the model that adapts only the spectrogram for the frontend feature.
 Donahue et al. proposed melody transcription from the musical mixture utilizing codified music representation \cite{castellon2021codified} derived from the hidden representation of JukeBox \cite{dhariwal2020jukebox}, which is originally proposed for musical audio generation. 
 
\section{Methods}
We compare four SSL models: 1) Wav2Vec2.0 \cite{baevski2020wav2vec}, 2) WavLM \cite{chen2022wavlm}, 3) MERT \cite{li2023mert}, and 4) MapMusic2Vec \cite{li2022map}.
Each model takes a raw waveform as input and employs convolutional layers for feature extraction along with 12 Transformer encoder layers. 
The output of each model consists of a 768-dimensional vector per frame.
We utilized these models as the frontend for each singing voice understanding task.

\begin{figure}[ht]
    \centering
    \includegraphics[width=\columnwidth]{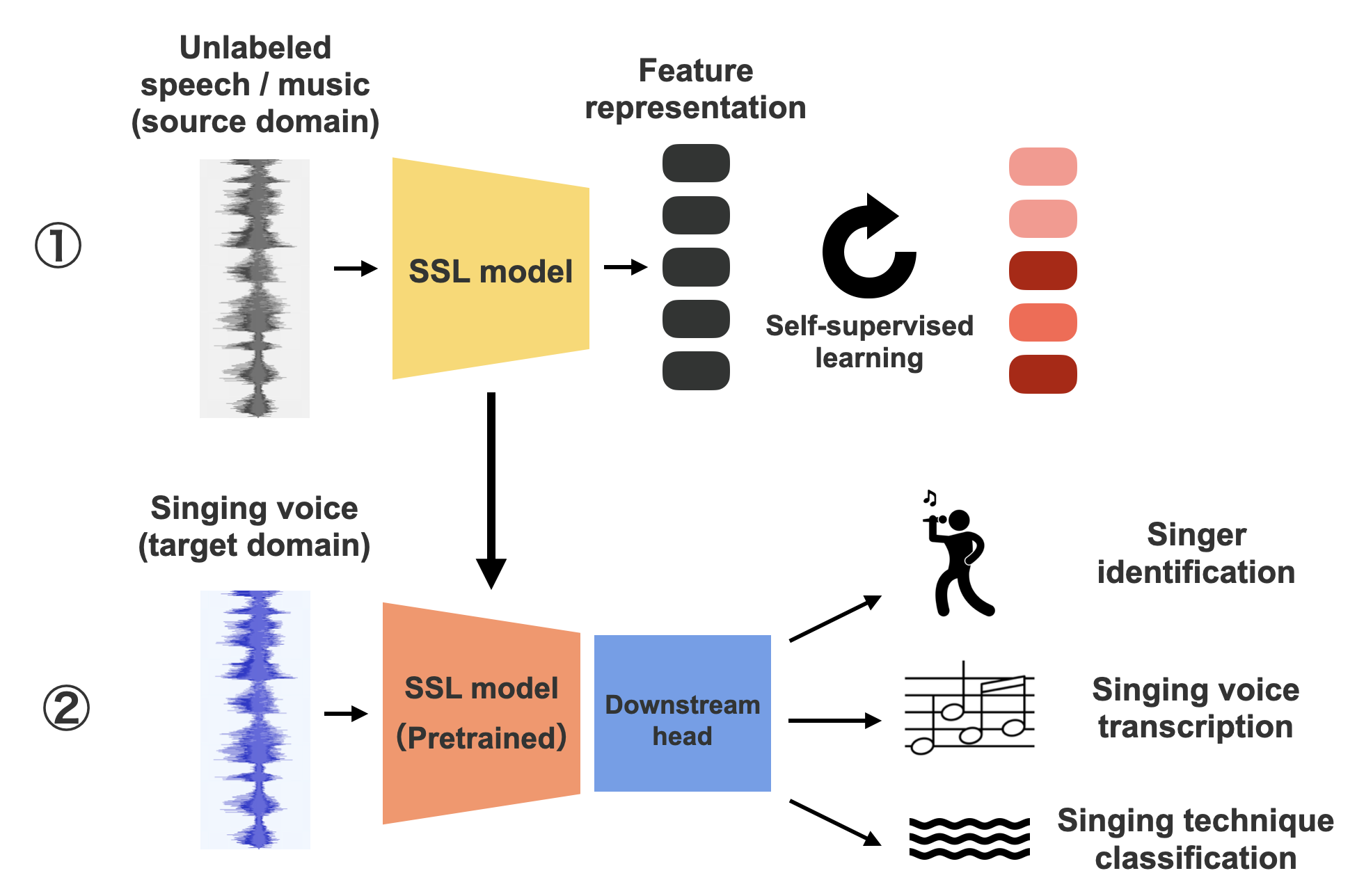}
    \caption{The concept of this paper. 1) Pre-training on upstream task: Utilizing a vast amount of  unlabeled data (either speech or music) and pre-training the model in a self-supervised fashion. 2) Transfer learning for downstream task: Leveraging pre-trained model and solving the target task (i.e., singing voice understanding tasks).}
    \label{fig:enter-label}
\end{figure}

\subsection{SSL models}
\subsubsection{Wav2Vec2.0}
Wav2Vec2.0 \cite{baevski2020wav2vec} is the model that has convolutional and Transformer Encoder layers and acquired speech representation through contrastive learning and masking.
It takes raw speech waveforms as input, and the initial convolutional layers produce latent representations denoted as $z$. To facilitate contrastive learning, the quantization module is applied to convert $z$ into discrete representations denoted as $Q$. Simultaneously, $z$ is fed into the Transformer layers after applying random masking to several frames. The output feature $C$ is then derived from the Transformer layers. Finally, contrast learning is performed masked time step in $C$ and $Q$. Here, the same time steps are considered as positive examples and different time steps are considered as negative examples. 
 Wav2Vec2.0 has shown its effectiveness in several downstream speech tasks by fine-tuning \cite{wang2021fine,vaessen2022fine,yi2020applying}. We used the Wav2Vec2.0 Base model\footnote{\url{https://huggingface.co/facebook/wav2vec2-base-960h}}.

\subsubsection{WavLM}
WavLM \cite{chen2022wavlm} is a large-scale pre-trained model with 94k hours of speech data as input that can treat full-stack speech processing.
It adopts masked prediction of hidden units like HuBERT \cite{hsu2021hubert} and denoising of the input speech as self-supervised pretraining.
The diversity of the pre-trained corpus of WavLM is wider than that of Wav2Vec2.0; Gigaspeech \cite{chen2021gigaspeech}, a collection of audiobooks, podcasts, and YouTube and VoxPopli\cite{wang2021voxpopuli}, a collection of European Parliament, in addition to Librispeech. 
We used the WavLM Base plus model, which demonstrates better performance than the Base model\footnote{\url{https://huggingface.co/microsoft/wavlm-base-plus}}.

\subsubsection{MERT}
MERT \cite{li2023mert} is a large-scale pre-trained model using unlabeled music data.
It is also inspired by masked prediction of hidden units as WavLM while it uses CQT spectrogram as its target in addition to quantized acoustic features with the purpose of enhancing the pitch representative power. 
It can achieve the parameter efficiency compared to the JukeMIR \cite{castellon2021codified}, the musical audio representation derived from JukeBox. 
We used the public-v0 model\footnote{\url{https://huggingface.co/m-a-p/MERT-v0-public}},  which is only trained on a public music dataset.

\subsubsection{MapMusic2Vec}
MapMusic2Vec\footnote{\url{https://huggingface.co/m-a-p/music2vec-v1}} \cite{li2022map} is a model that is pre-trained using BYOL \cite{grill2020bootstrap} with 1k hours of music data.
It relies on two neural networks that have the same structure as each other, reffed to the teacher model and the student model. The parameters of the teacher model are updated according to the exponential moving average (EMA) of the student model.
The student model takes partially masked audio raw waveform as input while the teacher model takes unmasked one and outputs the prediction of hidden outputs from the last $K$ layers of the teacher model. 

\subsection{Downstream model}
We describe the overview of how to use the SSL model and the downstream models in Figure 2.
\begin{figure}
    \centering
    \includegraphics[width=\columnwidth]{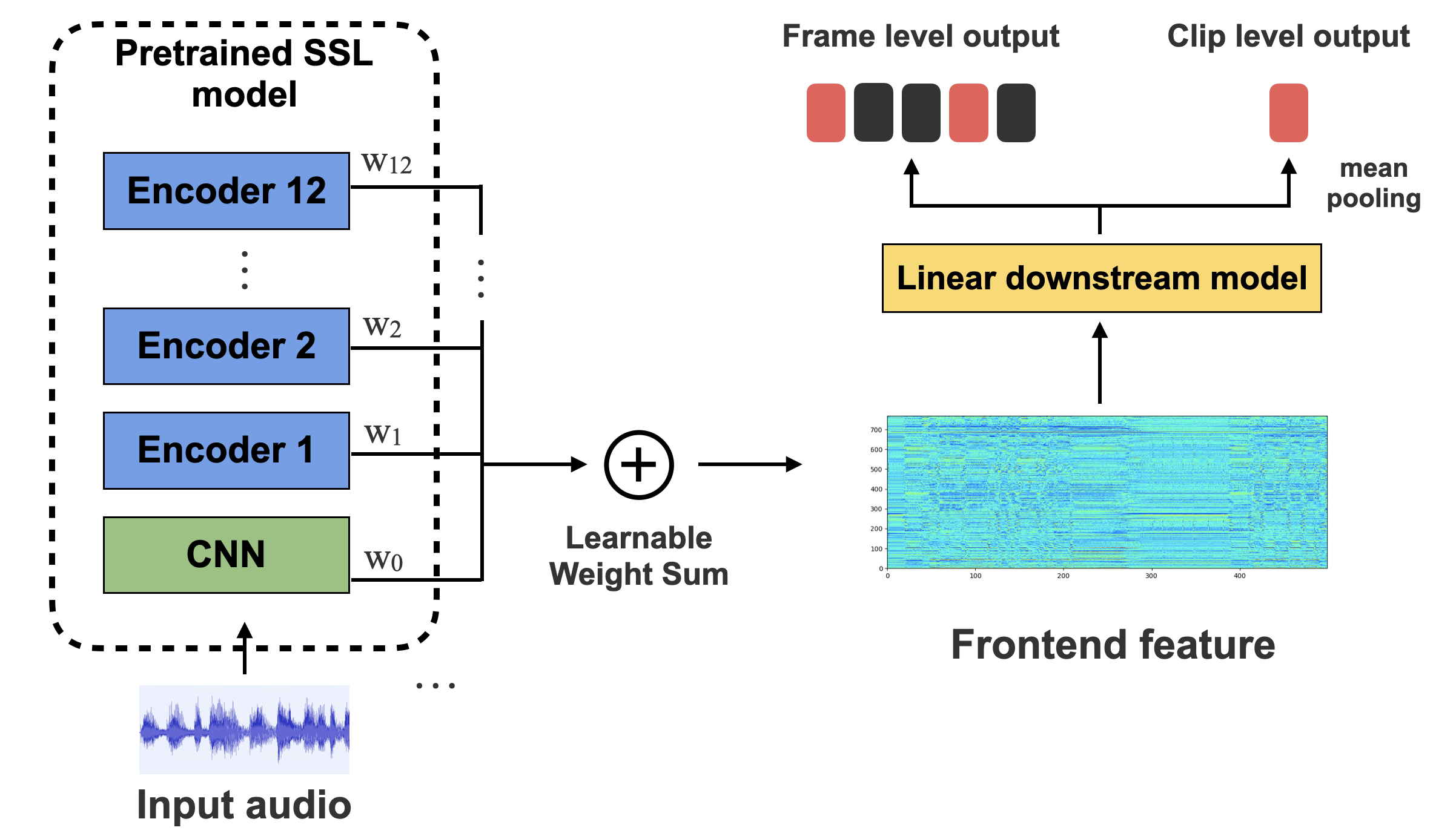}
    \caption{The overview of the whole model to solve the target tasks.}
    \label{fig:downstream}
\end{figure}
\subsubsection{Weighted sum}
We employed a weighted sum of the outputs from each Transformer encoder layer, including the input of the first layer, as the input for the downstream model. This approach was motivated by previous works (e.g.,\cite{choi2021neural,lin2022utility,fan2020exploring,pasad2021layer}) that have demonstrated different aspects of the input being captured by early, intermediate, and late layers of Transformer models. 
By using a weighted sum, we aimed to fully leverage the potential of SSL models.
We set 13 learnable weight values, each corresponding to the weight value assigned to the output of each layer.
\subsubsection{Classifier}
Given the aforementioned features, we use a linear downstream model to predict each.
The way of deriving the output of the target is depending on the tasks; For singing voice transcription, we directly used the frame-wise output. For the classification tasks (i.e., singer identification and singing technique classification), the feature is mean-pooled over the time axis to derive a clip-wise output.

\subsection{Fine-tuning}
In order to solve the downstream tasks, we fine-tune the models.
There are various strategies for fine-tuning large-scale pre-trained models, we follow the two-stage training as Gu et al. \cite{gu2023deep} did.
First, we freeze the parameter of the SSL models and make them learnable only on downstream models (i.e., the value of the weighted sum and the linear model). After several epochs, we unfreeze the Transformer encoders and fine-tune them. 

\section{Experiments and Results}
\subsection{Experiment on Singer Identification}
Singer identification is the classification task that identifies who is singing in a given sung audio clip.
\subsubsection{Experimental Condition}
We demonstrate 20-way singer identification using Artist20 \cite{ellis2007classifying} dataset, which collects 20 singers' music tracks.
The dataset contains six albums for each singer in the data set, for a total of 1,413 songs.
We split the dataset per album in order to avoid the leakage of production information about an album over the training and test set. We assign four albums for \textit{train}, one album for \textit{validation}, and the rest one album for \textit{test} subset.
Since the audio clips of the dataset include accompaniment of musical instruments, we applied vocal separation using Demucs V4 \cite{rouard2022hybrid}. 
Then, we split them into five-second chunks without overlapping at a sample rate of 16kHz and discarded non-vocal chunks by RMS filtering \cite{heydari2022singing}.

For training the models, we used Adam optimizer \cite{kingma2014adam} with 30 epochs.
We set the learning rate of $3 \times 10^{-3}$ for the first stage on the first six epochs and $5 \times 10^{-5}$ for the second stage with the remaining epochs. The batch size is set to 32.

We evaluated the models by the following metrics: F1-score, Top-2 accuracy, and Top-3 accuracy. 
For the baseline for the comparison, we adopt the \textbf{CRNN} model by Hsieh et al.\cite{hsieh2020addressing}. 
The model takes a 128-dimensional mel spectrogram as input and consists of four convolutional layers and two GRU \cite{cho2014properties} layers.
We re-implemented the model in order to measure Top-2 and Top-3 accuracy since \cite{hsieh2020addressing} only reported F1-score.
\subsubsection{Results} 
Table \ref{tab:sid} shows the results of singer identification. 
SSL models are demonstrating superior performance compared to the conventional SoTA model (i.e., CRNN) of singer identification. 
Notably, WavLM exhibited the highest performance in terms of F1 score, while MapMusic2Vec excelled in achieving the highest accuracy for Top-2 and Top-3 accuracy. 
These outcomes collectively suggest that the pre-training of SSL models with either music or speech data is leveraged in encoding the information associated with the singers.
\begin{table}[!ht]
    \centering
    \caption{The results of singer identification.}
    \begin{tabular}{l|ccc}
    \hline
        Methods & F1-score & Top-2  & Top-3 \\ \hline\hline
        Wav2Vec2.0 & 60.0 & 70.7 & 76.3 \\ 
        WavLM & \textbf{61.9} & 70.2 & 76.4 \\ 
        MERT & 56.8 & 68.4 & 75.6 \\ 
        MapMusic2Vec & 59.6 & \textbf{71.5} & \textbf{77.0} \\ \hline
        CRNN \cite{hsieh2020addressing} & 49.5 & 63.4 & 71.3 \\ \hline
    \end{tabular}
    \label{tab:sid}
\end{table}
\subsection{Experiment on Singing Voice Transcription}
Singing voice transcription refers to the process of converting sung audio signals into corresponding musical notes. In this paper, we employed the piano roll representation as the target for the singing voice transcription.

\subsubsection{Problem Definition}
We follow the settings of Wang et al. \cite{wang2021preparation}.
The target has four attributes: \textit{onset}, \textit{silence}, \textit{pitch class}, and \textit{octave}.
The beginning of silence is considered as the offset instead of a direct estimation of them due to its difficulty.
We set the pitch range from C2 (MIDI number 36, 65.41Hz) to B5 (MIDI number 83, 987.77Hz), therefore the target of \textit{octave} is four classes (i.e., 2-5.) 
In addition, the class of inactive is added on \textit{octave} and \textit{pitch class}, respectively.
Eventually, each frame contains 20-dimensional vectors as a target (i.e., \textit{onset} and \textit{silence} are binary, \textit{pitch class} is five classes, and \textit{octave} is 13 classes). 

\subsubsection{Experimental Condition}
We use MIR-ST500 \cite{wang2021preparation}, which consists of 500 Chinese pop songs with manually annotated vocal melody notes.
The authors of \cite{wang2021preparation} provide the official data split that allocates 400 songs for the training set and 100 songs for the test set.
Therefore, we followed the split as is, using the training set for the model training and the test set for the evaluation. 
Since the audio clips of the dataset include accompaniment of musical instruments, we applied vocal separation using Demucs V4 \cite{rouard2022hybrid}. 
During the training of the model, we split the input audio into five-second chunks without overlaps at a sample rate of 16kHz.
Given the input, the model outputs the prediction of the aforementioned 20-dimensional target 
 vector per frame. 
The frame length is about 20ms.
Suppose the prediction for \textit{onset}, \textit{silence}, \textit{pitch class} and \textit{octave} are $\hat{O}, \hat{S},\hat{P},\hat{V}$, the objective functions are as follows:
\begin{equation}
\begin{split}
\mathcal{L}_{svt} = \frac{1}{T} \sum_{t=1}^{T} \left[ BCE(\sigma(\hat{O}_t),O_t,w_o) + BCE(\sigma(\hat{S}_t),S_t,w_s) \right. \\
\left. + CE(\hat{P}_t,P_t) + CE(\hat{V}_t,V_t) \right]
\end{split}
\end{equation}
Where $T$ denotes the number of frames in the input, $\sigma(\cdot)$ denotes the sigmoid function, $BCE$ denotes binary cross-entropy loss, and $CE$ denotes cross-entropy loss.
Considering the imbalance between positive and negative samples, the weight is applied to the binary cross entropy loss. 
The values are $w_o=15.0$ for onset, and $w_s=1.0$ for silence, respectively.
For training the models, we used Adam optimizer \cite{kingma2014adam} with 30 epochs.
We set the learning rate of $3 \times 10^{-3}$ for the first stage and $5 \times 10^{-5}$ for 
the second stage.

We follow the strategy of postprocessing as Gu et al. \cite{gu2023deep} for deriving the actual estimation.
Briefly, each of the note properties is determined as follows:
\begin{itemize}
    \item onset: Setting a threshold value of 0.4. If the prediction value is higher than the threshold and the local maximum, the frame is set to onset.
    \item offset: $\argmin(\hat{S_t}>0.5)$ of the estimated silence sequence and after the onset time.
    \item pitch class and octave (i.e., midi number): Assign the mode of the estimated value between the onset and the offset time.
\end{itemize}
To evaluate the performance of our model, we adopted three evaluation metrics proposed in \cite{molina2014evaluation}, namely F1-score of COn (correct onset), COnP (correct onset and pitch), and COnPOff (correct onset, offset, and pitch). We utilized the mir\_eval library to calculate these metrics, using the default parameters: 50 cents for pitch tolerance, 50 ms for onset tolerance, and the larger value between 50 ms and 0.2 of the note duration for offset tolerance.

In order to establish baselines for comparison, we considered several conventional works:
\begin{enumerate}
\item \textbf{EfficientNet-b0}: This approach is based on utilizing the EfficientNet-b0 model \cite{tan2019efficientnet}, which was originally proposed as the baseline for the MIR-ST500 task \cite{wang2021preparation}.
\item \textbf{JDC$_{note}$}: This model was trained on pseudo-labeled data obtained through quantization of automatically detected vocal melody contours \cite{kum2022pseudo}.
\item \textbf{Wav2Vec2-Large}: This model employed the Wav2Vec2.0-Large model for the frontend and feeds only the last layer's output to the downstream model \cite{gu2023deep}.
\end{enumerate}

\subsubsection{Results}
Table \ref{tab:svt} shows the results of singing voice transcription.
MERT achieved the best score on COn and COnP among the four SSL models, while MapMusic2Vec demonstrated the best score on COnPOff.
We observed that the speech models (i.e., Wav2Vec2.0 and WavLM) exhibit lower COnP compared to the music models (i.e., MERT and MapMusic2Vec). It suggests that the music models have already acquired the representation related to musical notes while the speech models lack such incorporation due to the gap between speech and singing voice (i.e., the length of stable pitch region, musical expression, etc.).
In terms of performance comparisons with conventional works, every SSL model showed comparable performance with \cite{wang2021preparation} and \cite{kum2022pseudo}.
In addition, MERT outperforms Wav2Vec2Large in the COnP with fewer parameters.  
\begin{table}[!ht]
    \centering
    \caption{Results of singing voice transcription. All values are expressed in \%. The best-performing condition among all conditions is highlighted in bold, and the best-performing condition among SSL models is underlined.}
    \begin{tabular}{l|ccc}
    \hline
        Methods & COnPOff & COnP & COn \\ \hline \hline
        Wav2vec2.0 & 44.8 & 67.0 & 76.3 \\ 
        WavLM & 44.4 & 67.3 & 76.9 \\
        MERT & 46.7 & \underline{\textbf{71.6}} & \underline{78.2} \\ 
        MapMusic2Vec & \underline{50.7} & 70.0 & 77.9 \\  \hline
        EfficientNet-b0 \cite{wang2021preparation} & 45.8 & 66.6 & 75.4 \\ 
        JDC$_{note}$ \cite{kum2022pseudo}& 42.2 & 69.7 & 76.2 \\ 
        Wav2Vec2-Large \cite{gu2023deep}& \textbf{52.4} & 70.7 & \textbf{78.3} \\ \hline
    \end{tabular}
    \label{tab:svt}
\end{table}

\subsection{Experiment on Singing Technique Classification}
Singing technique classification is the task that identifies a singing technique that appeared in a given input audio clip. 
\subsubsection{Experimental condition}
We used VocalSet \cite{wilkins2018vocalset}, which is a publicly available dataset that annotated singing techniques.
VocalSet contains singing voices by 20 different professional singers (9 female and 11 male), performing 17 different singing techniques in various contexts.
We selected ten techniques (``belt,'' ``breathy,'' ``inhaled singing,'' ``lip trill,'' ``spoken excerpt,'' ``straight tone,'' ``trill,'' ``trillo,'' ``vibrato,'' and ``vocal fry'')  by all singers for the classification.
\begin{table}[hbt]
  \caption{Selected samples from VocalSet.}
  \label{tab:label_vocalset}
  \centering

  \begin{tabular}{l|cc}
  \hline
  Label name & Type of fluctuation & Samples \# \\
  \hline
  \hline
  straight & None & 1241\\
  belt & Timbre & 423\\
  breathy & Timbre & 455\\
  vocal fry & Timbre, Modulation & 587\\
  vibrato & Modulation & 1034\\
  trill & Modulation & 323\\
  trillo & Modulation & 242\\
  lip trill & Modulation & 376\\
  inhaled & Other & 151\\
  spoken & Other & 73\\
  \hline
  \end{tabular}
  
  \end{table}
We used the officially provided data split: 15 singers for the training set and 5 singers for the test set.
We trimmed the silence from the audio and split it into non-overlapping chunks of 3 seconds.
Typically the sampling rate of the audio tracks of VocalSet is 44.1kHz, the audio is resampled to 16kHz.
For training the models, we used Adam optimizer \cite{kingma2014adam} with 20 epochs.
We set the learning rate of $3 \times 10^{-3}$ for the first stage on the first five epochs and $5 \times 10^{-5}$ for the second stage with the remaining epochs. 
The batch size is set to 16. 

The data distribution of VocalSet is imbalanced over the classes that affect the classification performance \cite{yamamoto22_interspeech}.  Therefore, we adopt the inverse frequency weight for the loss function.
The typical weight value $w_c$ for the class $c$ is as follows: 
\begin{equation} \label{eq:alpha}
    w_c = \frac{1}{(n_c) ^ \alpha}
\end{equation}
where $n_c$ is the number of training samples in class $c$, and
$\alpha$ is the smoothing factor, which controls the smoothing of the loss weights.
Note that $\alpha$ = 0 corresponds to the value of 1 (i.e., no weighting) and $\alpha$ = 1 corresponds to a reciprocal number (i.e., weighting by the inverse class frequency). 
We set $\alpha$ = 0.2, which performed the best score in \cite{yamamoto22_interspeech}.
We evaluated the models by the following metrics: F1-score, Accuracy, Balanced accuracy, Top-2 accuracy, and Top-3 accuracy. 
We consider several conventional works as baselines.
\begin{enumerate}
    \item \textbf{1DCNN}: This model utilizes a CNN architecture that directly takes raw waveform as inputs. It serves as the official baseline model of the Vocalset dataset \cite{wilkins2018vocalset}.
    \item \textbf{OblongCNN}: This model employs a CNN architecture that takes a multi-resolution spectrogram, consisting of stacked representations with three different time-frequency resolutions as input. 
    Additionally, it incorporates four convolutional layers of varying shapes \cite{yamamoto2021investigating}.
    \item \textbf{D-CNN-cRT}: This model replaces the standard convolutional layers with Deformable convolution and employs Classifier Retraining (cRT) \cite{kangdecoupling} for training with a focus on addressing class imbalance \cite{yamamoto22_interspeech}.
\end{enumerate}
\subsubsection{Results}
Table \ref{tab:stc} shows the results of singing technique classification.
MapMusic2Vec exhibits the best performance in four SSL models and comparable performance to other conventional approaches.
It also achieved higher accuracy compared to the best-performing method among the conventional works, D-CNN-cRT.
\begin{table}[!ht]
    \centering
    \caption{The results of singing technique classification.}
    \begin{tabular}{l|ccccc}
    \hline
        Methods & F1 & Acc & BAcc & Top-2  & Top-3  \\ \hline \hline
        Wav2Vec2.0 & 56.1 & 58.3 & 60.1 & 74.8 & 82.1 \\ 
        WavLM & 55.6 & 60.8 & 57.9 & 75.3 & 83.8 \\ 
        MERT & 54.1 & 58.5 & 58.5 & 76.1 & 85.3 \\ 
        MapMusic2Vec & \underline{60.8} & \underline{\textbf{66.0}} & \underline{62.1} & \underline{79.0} & \underline{86.9} \\ \hline
        1DCNN \cite{wilkins2018vocalset} & 48.8 & 58.4 & 48.4 & 76.4 & 86.3 \\ 
        OblongCNN \cite{yamamoto2021investigating} & 51.3 & 55.4 & 57.5 & 74.3 & 85.8 \\ 
        D-CNN-cRT \cite{yamamoto22_interspeech} & \textbf{62.0} & 65.6 & \textbf{65.5} & \textbf{81.5} & \textbf{88.7} \\ \hline
    \end{tabular}
    \label{tab:stc}
\end{table}

\subsection{Layer-wise contribution analysis}
\begin{figure}[h]
    \centering
    \includegraphics[width=\columnwidth]{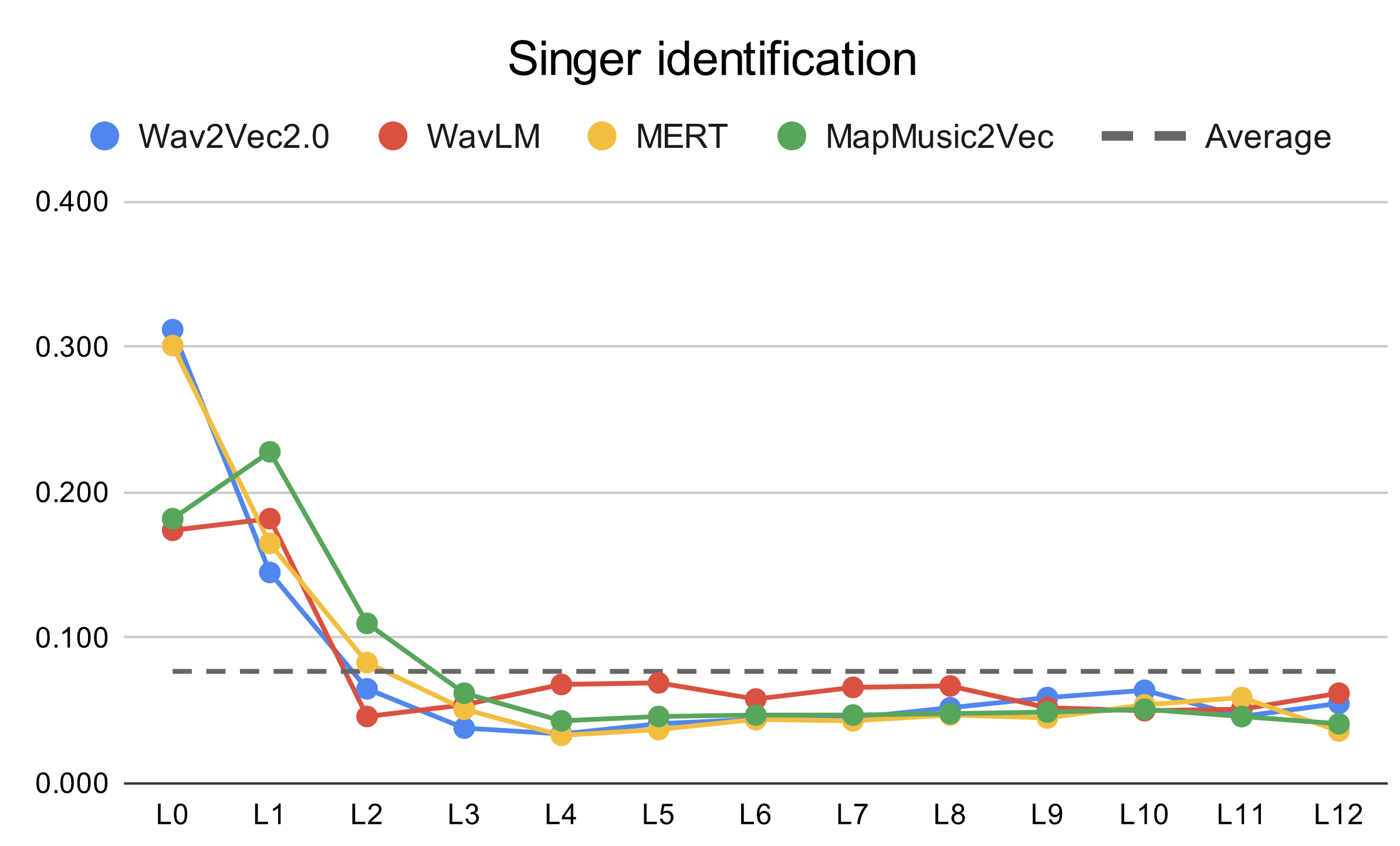}
    \caption{The weights after training on singer identification.}
    \label{fig:sid_lw}
\end{figure}

\begin{figure}[h]
    \centering
    \includegraphics[width=\columnwidth]{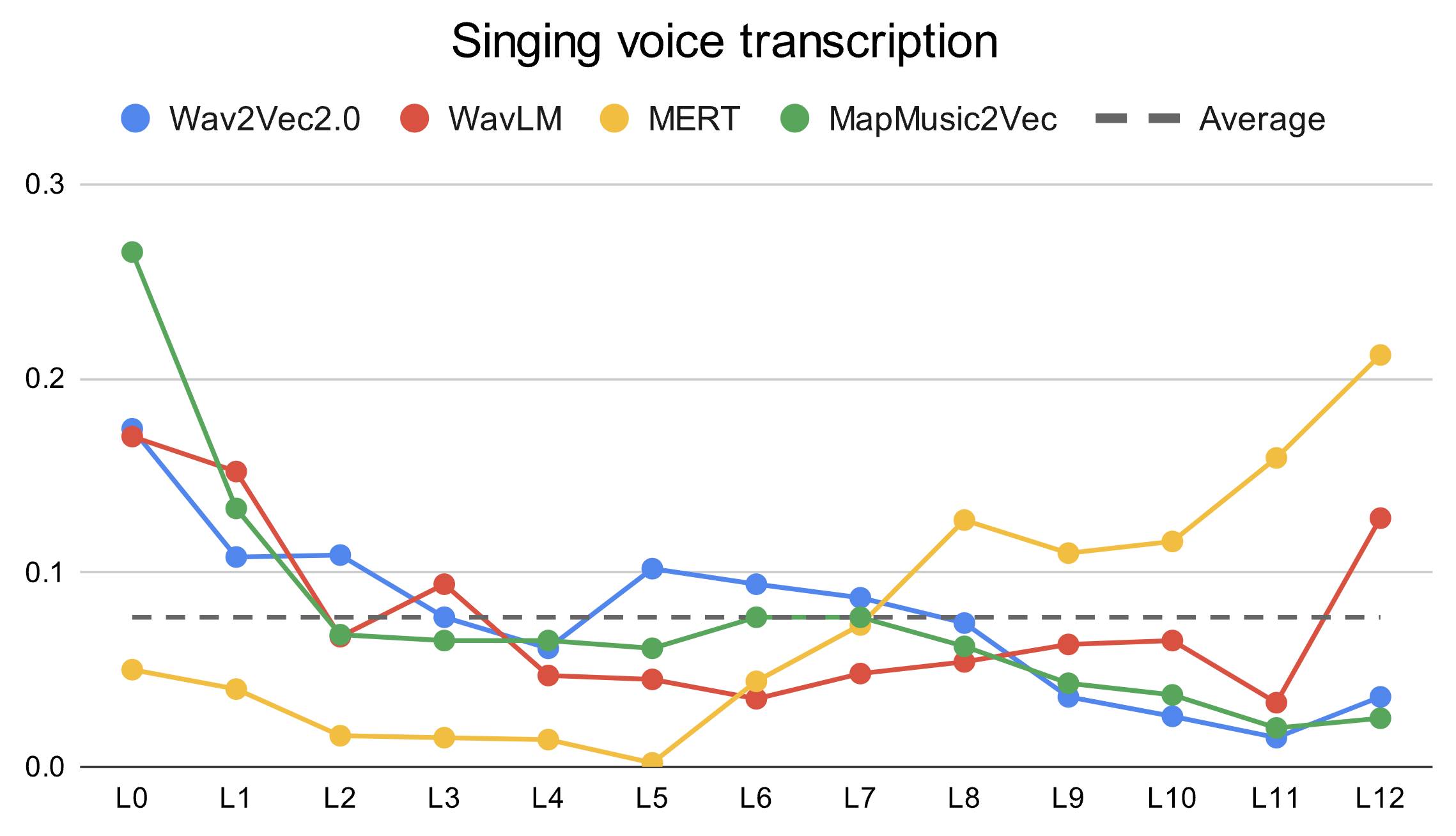}
    \caption{The weights after training on singing voice transcription.}
    \label{fig:svt:lw}
\end{figure}

\begin{figure}[h]
    \centering
    \includegraphics[width=\columnwidth]{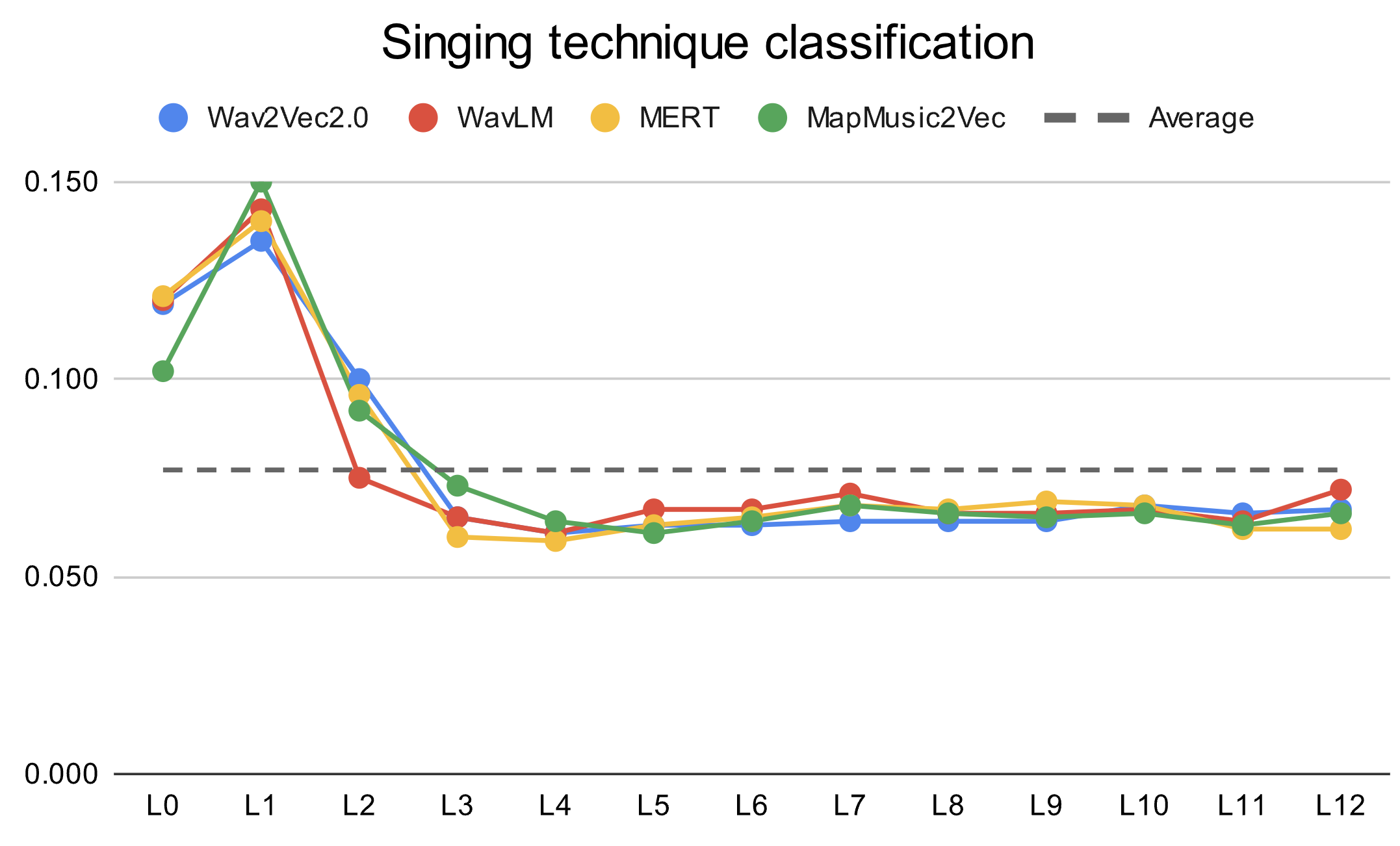}
    \caption{The weights after training on singing technique classification.}
    \label{fig:stc:lw}
\end{figure}

We further analyze the weight of each encoder layer's output for each SSL model.
Figure \ref{fig:sid_lw}, \ref{fig:svt:lw}, and \ref{fig:stc:lw} show the weights after training on singer identification, singing voice transcription, and singing technique classification, respectively. 
In each figure, `Ln' denotes the weight of the n-th encoder layer's output (i.e., L0 is for the input of the first layer.). 

For singer identification, the strong contribution lies in the early layers of each SSL model.
This finding aligns with previous research investigating the layer-wise contribution in various speech-related tasks, such as those presented in several works \cite{chen2022wavlm,chen2022large}. 
These studies suggest that speaker-related information tends to be captured in the early layers of the models.
Consequently, it can be inferred that a similar pattern holds true for singers, indicating that the crucial information for singer identification is also encoded in the early layers of the SSL models.

For singing voice transcription, except for MERT, the contribution also tends to lie in the early layers of each SSL model. 
According to \cite{lin2022utility}, low-level prosodic features such as pitch or loudness tend to locate the early layers of speech SSL models. 
Since pitch information is the most important for singing voice transcription, the results that align with the work are unsurprising.
In the case of MERT, CQT spectrogram is utilized as a prediction target.
Therefore, it is plausible that the last layers of MERT contain pitch-related information. 
As several works \cite{yong2023a,deng2022end} reported, singing voice transcription is relating the information of phonemes as well as the pitch (i.e., onsets are often accompanied by the transition of the lyrics). It might cause this "unsmooth-shaped" distribution compared to other tasks.

In the case of singing technique classification, the early layers exhibit greater values compared to other layers in each model.
Specifically, the first layer holds the greatest significance.
It might be because singing techniques are influenced by pitch modulation and timbre variation. 
However, there is also a possibility that it is affected by the data imbalance where vibrato and straight, whose characteristics are the status of pitch modulation, are the majority classes.


\section{Conclusion}
In this study, we address the challenge of limited data availability in singing voice understanding tasks by leveraging transfer learning of pre-trained self-supervised (SSL) models.
We employ four different SSL models as the frontend of our target task model. 
The models are subjected to comprehensive experiments encompassing singer identification, singing voice transcription, and singing technique classification.
Our experimental results demonstrate that each SSL model achieves comparable and in some cases superior, performance when compared to conventional state-of-the-art models. Additionally, we conduct layer-wise analysis to inspect the behaviors, specifically analyzing the weights of each layer.
Moving forward, our future research endeavors will focus on further investigating the impact of feature representation in addressing the data scarcity issue in automatic singing voice understanding tasks.
Furthermore, we propose that future studies can explore other singing voice understanding tasks, such as vocal melody extraction \cite{rao2022melody}, lyric transcription \cite{gao2022automatic}, and singer diarization \cite{suda2022singer}, among others, to expand the scope of research in this domain.

\section{Acknowledgements}
I am grateful to Hiroko Terasawa and Juhan Nam, for their fruitful comments on the methods and evaluation of models.
I also appreciate Jun-You Wang from National Taiwan University, for assisting in the preparation of \textit{MIR-ST500}.

This work was supported by JST SPRING, Grant Number JPMJSP2124.

\bibliography{r}
\bibliographystyle{IEEEtran}
\end{document}